\documentclass[10pt]{article}

\usepackage{times}
\usepackage{graphicx}
\usepackage{amsmath,amssymb}
\usepackage{multirow}
\usepackage{caption}
\usepackage{threeparttable}
\usepackage [latin1]{inputenc}

\newenvironment{sciabstract}{%
\begin{quote} }
{\end{quote}}

\newcounter{lastnote}

\title{Detection of Ultra High Energy Cosmic Rays and Neutrinos with Lunar Orbital Radio Telescope}

\author
{Linjie~Chen,$^{1,2\ast}$ Marc Klein Wolt,$^{3}$ Amin Aminaei,$^{4}$ Stijn Buitink,$^{5}$ Heino Falcke$^{3}$\\
\\
\footnotesize{$^{1}$National Space Science Center, Chinese Academy of Sciences, Beijing, 100190, China}\\
\footnotesize{$^{2}$National Astronomical Observatories, Chinese Academy of Sciences, Beijing, 100101, China}\\
\footnotesize{$^{3}$Astronomical Institute, Radboud University Nijmegen, Heijendaalseweg 135, 6525 AJ Nijmegen, The Netherlands}\\
\footnotesize{$^{4}$UC Davis, Dept. of Physics and Astronomy, Physics Bldg, 1 Shields Ave, Davis, USA}\\
\footnotesize{$^{5}$Astrophysical Institute, Vrije Universiteit Brussel, Pleinlaan 2, Brussels, 1050, Belgium}\\
\footnotesize{$^\ast$Corresponding author; E-mail:  ljchen@nao.cas.cn.}
}

\date{Jan. 25, 2023}

\begin{document}


\baselineskip18pt


\maketitle


\begin{sciabstract}
Particle cascades induced by ultra-high-energy (UHE) cosmic rays and neutrinos impacting on the lunar regolith usually radiate Cherenkov radio emissions due to the presence of excess negative charge, which is known as Askaryan effect. Several experiments have been carried out to detect the Cherenkov radio emissions in the lunar regolith. To prepare for future lunar Ultra-Long Wavelength (ULW, frequencies below 30 MHz) radio astronomy missions, we study the detection of the Cherenkov radio emissions with the ULW radio telescope that are operating at the lunar orbit. We have carried out instrument modelling and analytic calculations for the analysis of aperture, flux and event rate, and the analyses show the detectability of the Cherenkov radiation. Based on the properties of the Cherenkov radiation, we have demonstrated that the cosmic ray and neutrino events could be reconstructed with the three ULW vector antennas onboard the lunar satellites via measurements of the Askaryan radio pulse intensity, polarizations, etc. The results obtained by this study would be useful for future lunar radio explorer mission, where the detections of UHE cosmic rays and neutrinos could be successfully attempted.
\end{sciabstract}


\section{Introduction}           
\label{sec:intro}
Studies on the behaviour of ultra-high energy (UHE) particles and their source of origins have been a major topic of interest in modern high energy astrophysics and particle physics. However, the origin of ultra-high-energy (\(> 10^{18}\) eV) cosmic rays (UHECR) still remains unknown to date. It is mainly because of their deflected trajectories caused by cosmic magnetic fields. Above a threshold energy of \(\sim 10^{19.6}\) eV, cosmic rays usually interact with cosmic microwave background to produce pions and lose a substantial amount of energy within a distance of tens of mega parsec (Mpc). This is well known as Greisen-Zatsepin-Kuzmin (GZK) effect (\cite{greisen1966},~\cite{zatsepin1966}). The few cosmic rays that observed above this threshold energy must originate from the distant sources. In order to determine the source origin of such UHE cosmic rays, the detection of UHE neutrinos, those produced together with the cosmic rays or during the GZK interaction, has usually been employed as one of the method. Since neutrinos are chargeless and weakly interacting particles, the propagating UHE neutrinos (UHEC\(\nu\)) usually remain unaffected over large cosmic distances, and therefore, their arrival directions carry direct information of their source of origins.

The chief problem of both UHECR and UHEC\(\nu\) event detection come from their extreme rarity. The flux at GZK energy is around one particle per square kilometer per century, which calls for huge detectors (thousands km\(^{2}\)) to collect a significant amount of events at ultra-high energies. For instance, even the Pierre Auger Observatory (PAO) with a collecting area of 3000 km\(^{2}\) can only detect of order 30 particles above the GZK energy per year \cite{james2019}. However, for the detection of particles above \(10^{21}\) eV, the detectors are needed to have a thousand-fold increase in their collecting area. Thus, space-based detection systems are, therefore, appeared to be a feasible approach that is currently being attempted and implemented.

The Moon acts as a huge natural detector with large collecting area and is first proposed to be a target to detect cosmic ray and neutrino showers in \cite{dagkesamanskii1989}. The detection is based on the Askaryan effect, which is pointed out by G.A. Askaryan from the Lebedev Physical Institute in 1961 \cite{askaryan1962}. When high-energy particle interaction occurs in a denser medium like ice, rock salt, lunar regolith, or the atmosphere, it attracts electrons from the surrounding medium and transfer them into the shower disk. With the annihilation of showering positrons in flight, a net excess of electrons are produced. The excess electrons rapidly appear and disappear to produce a radio pulse that could be observed by a sensitive telescope. Using the above technique several experiments have been proposed and performed to detect high-energy cosmic particles, such as Parkes (\cite{hankins1996},~\cite{hankins2001}), GLUE (\cite{gorham2000},\cite{gorham2004}), LUNASKA \cite{james2010},\cite{james2009}, NuMoon \cite{buitink2010}, RESUN \cite{jaeger2010}, LaLuna \cite{spencer2010}, LOFAR \cite{singh2012}, FAST \cite{james2019}, all of which employ the terrestrial telescopes to carry out the observations. In the literatures (\cite{gusev2006},~\cite{stal2007},~\cite{aminaei2018}), the scheme of a lunar orbital satellite system was presented and analyzed to detect the UHE cosmic rays and neutrinos.

The ultra-long wavelength (ULW) (\(\geq 10~\ \)m) range, the last unexplored region of the electromagnetic (EM) spectrum, is presently one of the new promising areas of radio astronomy. Strengthening the science cases for extending the radio astronomy into the unexplored ULW window has been significantly increasing. Due to the reflection of the ULW signals by the Earth's ionosphere and the presence of the strong terrestrial anthropogenic radio frequency interference (RFI), the far side of Moon can be an ideal site for carying out the ULW radio observations. To date, various Moon-based telescopes have been made and launched for exploring the ULW region, such as OLFAR \cite{engelen2011}, FARSIDE \cite{mimoun2011}, DARE \cite{burns2012}, LRX \cite{marc2012}, DSL (\cite{boonstra2016},~\cite{chen2021}), NCLE \cite{boonstra2017}, LFRS \cite{ji2017}, etc. The lunar ULW radio telescopes provide ample opportunities to detect the Cherenkov radiation usually caused by the UHNCR or UHEC\(\nu\) on the Moon.

In this paper we present the radio detection of the UHNCR or UHEC\(\nu\) with the lunar orbital ULW radio antennas. The paper is organized as follows. In Section 2 we describe the model for the detection of the lunar Cherenkov emission. The aperture and flux limit for the lunar UHECR and UHEC events are estimated and analyzed in Section 3. In Section 4 we discuss the characteristic of the UHNCR or UHEC\(\nu\) with the multi-satellite detections. Finally, summary and conclusions are provided in Section 5.


\section{Model for radio emission detection}
\label{sec:model}
\begin{figure}[h]
\centering
\includegraphics[width=4.8in]{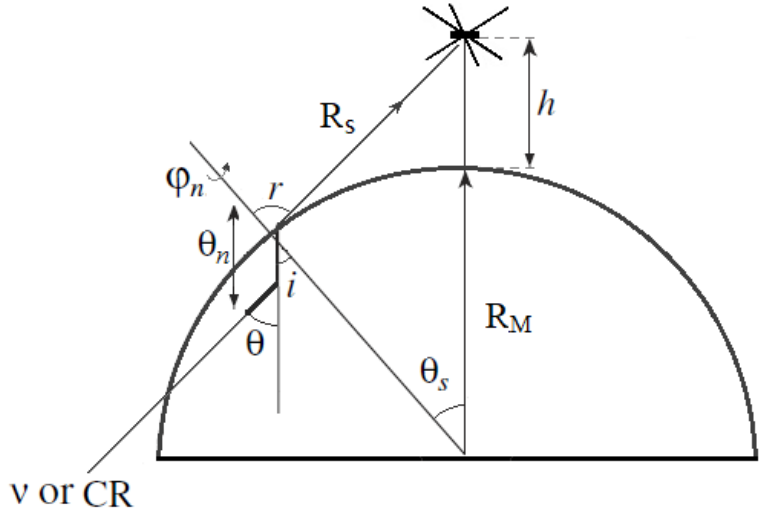}
\caption{Schematic view of lunar antenna detections of Askaryan radio emissions produced by the UHECR or UHEC\(\nu\) in the lunar soil. \label{fig:scheme}}
\end{figure}

When energetic particles interact in the denser medium like lunar regolith, intense coherent radio pulses are produced in the frequency range from MHz to GHz. The radio emissions that escaped from the lunar surface could be detected by a Moon-based or terrestrial radio telescope as shown in Figure \ref{fig:scheme}, which strongly depends on the sensitivity of the telescope. Though the terrestrial radio telescope have a huge collecting area and the best sensitivity, the intensity of the lunar Askaryan radio emission is weak due to the long-distance attenuation. A lunar radio telescope may provide a reasonable detection capability for the UHECR and UHEC\(\nu\) events since it is thousand times closer to the lunar surface than to the ground-based telescopes. For the detection of the UHE particles, the effective aperture is another vital factor which is directly dependent on the physical lunar surface illuminated by the antenna. The physical surface area is determined by the antenna beamwidth and antenna altitude from the lunar surface, and the altitude affects the detected electric field of the Askaryan radio emission. In order to study the lunar UHECR and UHEC\(\nu\) detections, the electromagnetic properties of Cherenkov radiation have been modelled including the system parameters of the radio detector.

\subsection{Lunar Cherenkov radio emission}
According to numerous Monte Carlo simulations, the Cherenkov radio emissions from the UHE hadronic showers have been parameterized for different dielectric media (\cite{zas1992},~\cite{alvarez1997},~\cite{alvarez2001},~\cite{gorham2004}). In the literatures (\cite{gusev2006},~\cite{stal2007},~\cite{gayley2009},~\cite{bray2016},~\cite{aminaei2018}) the radio detection of lunar Askaryan pulse caused by the UHECR and UHEC\(\nu\) interacting with Moon have been addressed. For the Cherenkov radiation, we utilized the formula given in Ref. (\cite{james2009},~\cite{gayley2009},~\cite{bray2016}) to calculate the electric field strength \(\varepsilon_{0}\) (\(\mu\)V~m\(^{-1}\)~MHz\(^{-1}\)) at the Cherenkov angle \(\theta_{\rm c} = \cos^{-1}(1/n)\).
\begin{eqnarray}
\label{eqn:intensity}
\boldsymbol{\varepsilon_{0}}(E, R, \nu) = 8.45\times10^{6}\frac{1}{R}(\frac{E_{\rm s}}{E_{0}})(\frac{\nu}{\rm GHz})(\frac{1}{1+(\nu/\nu_{0})^{1.23}})
\end{eqnarray}
where, \emph{R} is the distance from the source point to the detector, \(E_{\rm s}\) is the shower energy in eV. \(E_{0} = 10^{20} \rm eV, \nu_{0} = 2.32 \rm GHz\) (\cite{james2009},~\cite{bray2016}). The angular distribution of the radio emissions around the Cherenkov angle \(\theta_{\rm c}\) is crucial for the detection of lunar UHNCR and UHEC\(\nu\) events. Due to the loss of coherence, the radio emission decreases in the direction away from \(\theta_{\rm c}\). \cite{scholten2006} discussed the angular spread of the Cherenkov radiation around the Cherenkov angle. Through comparing several different formulae of intensity distribution, Scholten concluded that the gaussian parametrization formula is more accurate since it agrees with both the Monte Carlo simulations at high frequencies and the analytic results at low frequencies. In the present study, we use the formula given in \cite{scholten2006} for angular spread of the Cherenkov radiation as follows,
\begin{eqnarray}
\label{eqn:angular}
\hspace{-8mm}\boldsymbol{I}_{\theta} = \frac{\sin\theta}{\sin\theta_{\rm c}} e^{-\frac{Z^{2}}{2}}, ~~~Z=\frac{\cos\theta-\cos\theta_{\rm c}}{\Delta_{\rm c}\sin\theta_{\rm c}}
\end{eqnarray}

The spreading of the radiation intensity is determined by \(\Delta_{\rm c} = 0.0754[L(E_{0})/(\nu L(E_{s})]\) (in radians), which is inversely proportional to the shower length and the radiation frequency. Here \(L(E\) is the shower length determined by the energy, \(L(E) = 12.7+2/3\log(E/E_{0})\) \cite{scholten2006}.

In the detection analysis of lunar UHE particle events, a point-source is treated for the coherent Cherenkov radiation. When the Cherenkov radio emissions escape from the lunar surface, it will be reflected and transmitted at the interface. According the Fresnel law, the transmission coefficient for parallel polarization can be expressed as
\begin{eqnarray}
\label{eqn:transcoeff}
\hspace{-8mm}\boldsymbol{\hat{t}_{\|}(\theta_{s})} = \frac{2\cos r}{n\cos r + \cos i}
\end{eqnarray}
where \(\theta_{s}\) is the polar angle which is uniquely related to the angle \emph{r} and \emph{i}. \emph{r} is the angle of refraction relative to the normal (outside the Moon) as the rays pass through the lunar surface into free space, \emph{i} is the angle of incidence as shown in Figure \ref{fig:scheme}. \emph{r} and \emph{i} have the relation as \(\sin(i)=\sin(r)/n\), \emph{n} is the refraction index of lunar regolith, and chosen to be 1.73. The detected radiation from a lunar regolith shower, with energy \(E\), at a frequency \(\nu\) and an angle \(\theta\), can be expressed as \(\boldsymbol{\varepsilon}=\boldsymbol{\varepsilon_{0}}(E, R, \nu) \cdot \hat{t}_{\|}(\theta_{s}) \cdot I(\theta)\). Here the absorption of the radio waves before exiting the lunar surface has not been discussed, and it will be taken into account in the aperture calculations.
\subsection{Detection with lunar ULW radio antenna}
The radio emission from the hadronic cascades induced by the UHE cosmic rays or neutrinos in the lunar regolith covers a broad frequency band from MHz to GHz, and it peaks in the GHz regime \cite{james2009}. Therefore, most of the past lunar radio experiments to UHECR or UHEC\(\nu\) operate at GHz frequency band (\cite{hankins1996},~\cite{gorham2000},~\cite{james2010},~\cite{jaeger2010}). With the decreasing frequency the peak intensity of the radiation decreases, however, the increasing angular spread allows the radio emission detected by the lunar radio telescope to escape from the lunar surface within a broader solid angle, which makes it increasingly efficient to detect the Cherenkov radiation at low frequencies. Furthermore, the surface roughness does not play an important role in the detection analysis at lower frequencies since a considerable fraction of the radiation will penetrate the surface. According to these factors, some experiments running at low frequencies(\(\sim150\) MHz) for the lunar particle detection have been carried out or planned with the Westerbork Synthesis Radio Telescope (WSRT), NuMoon \cite{buitink2010}, the LOw Frequency ARray (LOFAR), and the Square Kilometer Array (SKA-low). Although the ULW band \(\leq 30 \rm MHz\) is beyond the optimum frequency window where the wavelength is comparable with the shower length \cite{scholten2006}, and the bandwidth is also limited at ULW band, the increased particle energy could well compensate the loss of electric field strength at lower frequency. The net effect is that the detection probability could increase reasonably at adequately high shower energies.

In the present study, we discuss the UHECR and UHEC\(\nu\) detection with lunar orbital ULW radio telescope. For the space ULW radio observations, a tripole antenna has been selected as the receiving element in most of the radio missions as DSL (\cite{boonstra2016},~\cite{chen2021}), NCLE \cite{boonstra2017}, LFRS \cite{ji2017}, and OLFAR \cite{bentum2009}, since it has specific advantages of measuring the three-dimension electric field, estimating the directions of arrival (DOA) of incident signals with single antenna (\cite{chen2010},~\cite{chen2018}), and protecting the desired signals from interference signals \cite{compton1981}. In the lunar ULW mission, the Dark Ages is undoubtedly one of the greatly interesting sciences, it requires the observation of the entire \(1-30\) MHz band, which could be extend up to 50 MHz for cross-referencing with the terrestrial radio facilities such as LOFAR. Therefore, the tripole antenna of the lunar ULW radio telescope is supposed to operate at the frequency range up to 40 or 50 MHz (\cite{ji2017},~\cite{boonstra2016}).

For a radio telescope, its sensitivity can be determined by the system equivalent flux density (SEFD)~\(F = kT_{sys}/A_{eff}\) in Jansky (\(10^{-26} ~\rm W \rm m^{-2} \rm Hz^{-1}\)), . Here \emph{k} is the Boltzmann's constant, \(\rm T_{sys}\) the system temperature and \(\rm A_{eff}\) the effective collecting area. However, a electric field strength (V/m/Hz) of a coherent radio pulse is usually utilized to characterize the Cherenkov radiation. In order to analyze the sensitivity of a radio telescope for the Askaryan pulse detection, it is required to express the telescope sensitivity in terms of the electric field strength. If a radio telescope has a flat system noise spectrum over the observed bandwidth, the equivalent root mean square (RMS) spectral electric field \cite{bray2016} can be written as follows,
\begin{eqnarray}
\label{eqn:efield}
\boldsymbol{\varepsilon_{\rm rms}} = \frac{\rm E_{\rm rms}}{\Delta\nu},~~~\rm E_{\rm rms} = \sqrt{\frac{kT_{sys}Z_{0}\Delta\nu}{A_{eff}}}
\end{eqnarray}
where \(\rm E_{\rm rms}\) is the RMS electric field strength over the bandwidth \(\Delta\nu\) in one polarization, and \(Z_{0}\) is the impedance of free space. Then the sensitivity of an antenna to detect a coherent radio pulse (the minimum detectable electric field) can be defined as \(\boldsymbol{\varepsilon_{\rm min}} = N_{\sigma}\varepsilon_{\rm rms}\), where \(N_{\sigma}\) is the minimum number of standard deviations needed to reject statistical noise pulses.

At the ULW band, the galactic synchrotron emission dominates the sky radio foreground, its spectrum varies greatly with the frequencies, the brightness temperature rises from about \(10^{4}\) K at 30 MHz to about \(2.6\times10^{7}\) K at around 1 MHz; Furthermore, the antenna effective collecting area changes significantly with the frequencies too. Therefore, the Equation (\ref{eqn:efield}) cannot be applied directly to the electric field calculation for a ULW radio antenna. In order to truly characterize the equivalent spectral electric field of the telescope, the RMS electric field strength \(\rm E_{\rm rms}\) should be re-computed by integrating the SEFD over the frequency. According to the studies in Ref. \cite{chen2018}, the sky noise temperature is absolutely dominant in the total system temperature for the ULW radio antennas of different lengths, then \(\rm E_{\rm rms}\) can be re-written as,
\begin{eqnarray}
\label{eqn:spectralefield}
\rm \boldsymbol{E}_{\rm rms} = \sqrt{kZ_{0}\int_{\nu_{\rm L}}^{\nu_{\rm H}}\frac{T_{sky}(\nu)}{A_{eff}(\nu)}d\nu}
\end{eqnarray}
\begin{table}[h]
\begin{center}
\caption[]{Sensitivity parameters of radio UHECR and UHEC\(\nu\) observations, Frequencies and bandwidths are in MHz.}
\label{tab:sensitivity}
\begin{tabular}{ccccccc}
\hline\noalign{\smallskip}
Simulations & \(\nu_{\rm min}\) & \(\nu_{\rm max}\) & \(\Delta\nu\) & \(\nu_{0}\) & \(\boldsymbol{\varepsilon_{\rm min}}\)(uV/m/MHz) \\
\hline
Case I & 10 & 30 & 20 & 20 & 2.06 \\
Case II & 15 & 25 & 10 & 20 & 2.84 \\
Case III & 10 & 40 & 30 & 25 & 1.61 \\
Case IV & 15 & 35 & 20 & 25 & 1.93 \\
Case V & 20 & 30 & 10 & 25 & 2.67 \\
Case VI & 10 & 50 & 40 & 30 & 1.34 \\
Case VII & 15 & 45 & 30 & 30 & 1.53 \\
Case VIII & 20 & 40 & 20 & 30 & 1.83 \\
Case IX & 25 & 35 & 10 & 30 & 2.55 \\
\noalign{\smallskip}\hline
\end{tabular}
\end{center}
\end{table}
Where \(\nu_{\rm L}\) and \(\nu_{\rm H}\) are the lower and upper limit of the frequency respectively. Using a sky temperature model \cite{heino2009}, the sensitivity of a ULW radio antenna longer than 5 meters \cite{chen2018} to detect the Askaryan radio pulse has been calculated with a nominal value \(N_{\sigma} = 3\) for different center frequencies and bandwidths, as shown in Table \ref{tab:sensitivity}. Note that the antenna length has less influence on the sensitivity since the sky noise temperature is absolutely dominant.

In order to improve the detection probability of lunar UHECR and UHEC\(\nu\) events, we use the central frequency of 30 MHz and the bandwidth of 40 MHz to obtain the best sensitivity in our analysis.

\section{Detection analysis for lunar UHECR and UHEC\(\nu\) events}
\label{sec:detection}
We used the method presented in Ref. \cite{gusev2006} to analyze the detection of lunar UHECR and UHEC\(\nu\) events. In order to decide the cascade event rates for the UHECR and UHEC\(\nu\), one way is in term of the aperture size, which is multiplied by the isotropic cosmic ray or neutrino flux in the given energy bin to obtain the detection rate. We calculate the analytic aperture by integrating the angular aperture for specified energy \emph{E} over the angles \((\theta_{s},\varphi_{s}\)) at the lunar surface \emph{S}, and over the angles \((\theta_{n},\varphi_{n}\)) of the cascade axis relative to the normal to this surface, as shown in Figure \ref{fig:scheme}.

For a volume element near the lunar surface \((z\simeq 0)\) with the polar angle \(\theta_{s}\), (see Figure ~\ref{fig:scheme}) radio emissions from it received by the detector has a given refraction angle \emph{r}, \(\sin(r)=\sin(\theta_{s})(R_{\rm M}+h)/R(\theta_{s})\). Here \(R(\theta_{s})=[R_{\rm M}^{2}+(R_{\rm M}+h)^{2}-2R_{\rm M}(R_{\rm M}+h)\cos\theta_{s}]^{1/2}\), is the distance between the surface element and the detector. The maximum value of the polar angle \(\theta_{s}^{max}=\arccos[R_{\rm M}/(R_{\rm M}+h)]\) is associated with the refraction angle \(\emph{r}\longrightarrow \pi/2\).

At low frequencies, the lunar surface roughness will not have a major influence on the detection sensitivity, and thus can be neglected \cite{scholten2006}, which could be another advantage of UHE particle detection with a lunar ULW radio antenna. In our analysis, the Moon is assumed to be an approximately smooth sphere.

\subsection{UHE cosmic rays}
For the UHE cosmic rays, they cannot penetrate through the lunar regolith, and cosmic ray events occur in the lunar regolith near the surface. Therefore, only the upper hemisphere (\(\cos\theta_{n}>0\)) contributes to the aperture calculation for the cosmic ray cascades. The total aperture \(A_{CR}(E)\) is defined by Equation (\ref{eqn:craperture}) and (\ref{eqn:crsolidangle}) as the integral of the angular aperture \(\Delta\Omega_{CR}(E,\theta_{s})\) over the surface S \cite{gusev2006}.
\begin{equation}
\label{eqn:craperture}
A_{CR}(E)=A_{0}\int\Delta\Omega_{CR}(E,\theta_{s})d\cos\theta_{s}
\end{equation}
\begin{equation}
\label{eqn:crsolidangle}
\begin{split}
\Delta\Omega_{CR}=\int\cos\theta_{n}\Theta [\boldsymbol{\varepsilon}(E,\theta_{s},\theta)-\boldsymbol{\varepsilon_{\rm rms}}]\times\Theta(\cos\theta_{n})d\varphi_{n}d\cos\theta_{n}
\end{split}
\end{equation}
where, \(A_{0} = 2\pi R_{\rm M}^{2}\) is the physical area of lunar hemisphere, the unit step function \(\Theta(\boldsymbol{\varepsilon-\varepsilon_{\rm rms})}\) only selects the cascade events that produce radio signals above the threshold value \(\boldsymbol{\varepsilon_{\rm min}}\). The radio attenuation length of lunar soil depends on the fraction of its composition. For the lunar regolith we use the value \(\ell=60\lambda\) (\(\lambda\), the wavelength in the vacuum) given in the literatures (\cite{james2009},~\cite{bray2016}), it is a quite large values at low frequencies. Thus the attenuation of radio wave prior to their escape from the lunar surface can be ignored since the cosmic ray cascades exist only few meters deep.
\begin{figure}[h]
\centering
\includegraphics[width=4.0in]{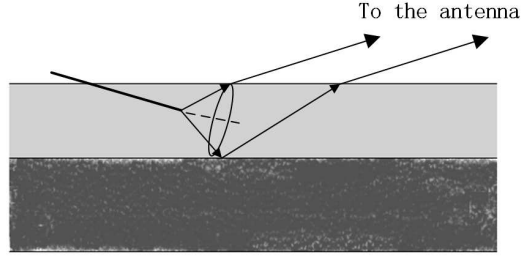}
\caption{Askaryan radio emissions initiated by UHE cosmic rays propagate between the lunar soil and lunar orbital radio antenna. The lunar soil is modelled with two layers, the regolith and the sub-regolith.
\label{fig:lunarsoilcr}}
\end{figure}

The lunar subsurface soil consists of several different strata according to the detection of Lunar Penetrating Radar. Therefore, the reflected radio waves by the strata interfaces that can be detected by the lunar orbital radio antenna should also be taken into account besides the direct radio waves emitted by the hadronic shower, as shown in Figure \ref{fig:lunarsoilcr}. Due to the very small contribution, the secondary reflections at the interfaces are neglected. In order to simplify the analysis, the lunar soil structure is modelled as only two layers, lunar regolith and lunar sub-regolith. The regolith is assumed to be 12 meter deep \cite{li2020}, and the refractive index of lunar regolith (\emph{n} = 1.73) and sub-regolith (\emph{n} = 2.5) (\cite{james2009},~\cite{bray2016}) have been used in the aperture calculation.
\begin{figure}[h]
\centering
\includegraphics[width=4.8in]{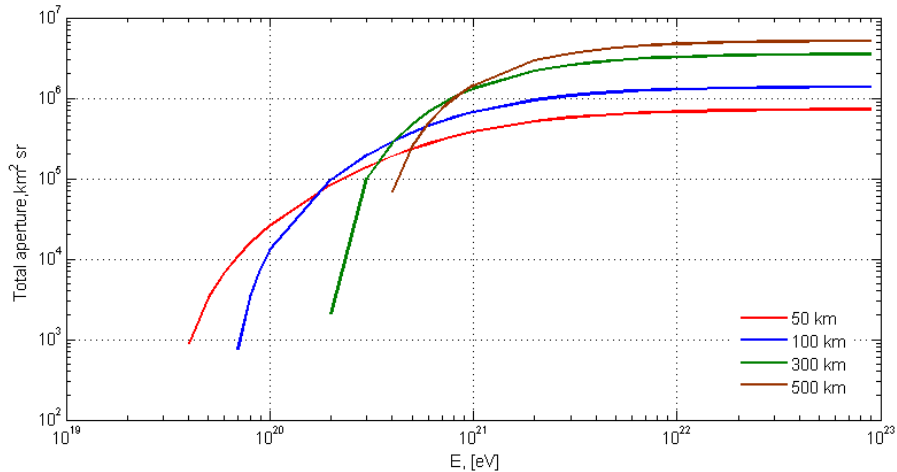}
\caption{Total aperture of UHE cosmic rays for a lunar orbital ULW radio antenna at different altitudes of 50 km, 100 km, 300 km, and 500 km with a sensitivity of 1.34 uV/m/MHz.
\label{fig:craperture}}
\end{figure}

Using analytic integrations of the angular aperture over the Moon surface, the total aperture are computed for the lunar orbital ULW radio antenna with different altitudes. As shown in Figure \ref{fig:craperture}, with the higher altitude of lunar ULW radio antenna, the physical aperture illuminated by the antenna increases, however, the detected Cherenkov radiation on the antenna becomes weaker due to the distance-decay effect. As a result, the minimum (threshold) cascade energy at which the radio emission intensity on the antenna is equal to the threshold intensity \(\boldsymbol{\varepsilon_{\rm min}}\) increases when the altitude rises. For all the altitudes, the aperture reaches around the peak above the energy of \(\sim10^{22}\) eV.

Based on the model-independent estimate \cite{lehtinen2004}, for the zero detected events, the upper limit for differential (in energy) flux \(J(E)\) of primary particles in the observation time \emph{T}, can be obtained by the following relation,
\begin{eqnarray}
\frac{dJ(E)}{dE} \leq \frac{S_{up}}{EA(E)T}
\end{eqnarray}
Here the Poisson factor \(S_{up}=2.3\) for a limit with \(90\%\) confidence level.
\begin{figure}[h]
\centering
\includegraphics[width=4.8in]{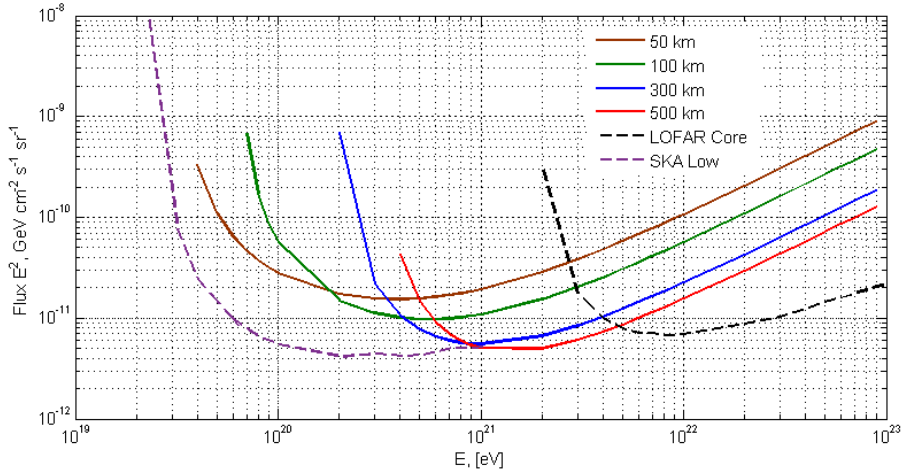}
\caption{UHE cosmic ray flux of a lunar orbital ULW radio antenna at different altitudes of 50 km, 100 km, 300 km, and 500 km for one-year observation. The results are compared with the flux limits of both SKA Low for a 1000-hour observation between 100 MHz and 350 MHz (reproduced from Bray et al., 2014) and LOFAR for a 90-day observation between 125 MHz and 175 MHz (reproduced from Singh et al., 2009).
\label{fig:crflux}}
\end{figure}

For one-year observation with the lunar orbital ULW radio antenna, the flux limits are calculated for different antenna altitudes. As shown in Figure \ref{fig:crflux}, only the cosmic ray with the energy above a threshold can be detected by the lunar orbital antenna, it is \(4\times10^{19}\) eV for the antenna at the 50 km high orbit, and \(4\times10^{20}\) eV for the 500 km high orbit. The results in Figure \ref{fig:crflux} also show that the flux limit of single lunar orbital antenna will be competitive with the estimated limits of large terrestrial radio telescopes such as LOFAR \cite{singh2009} and SKA-Low \cite{bray2014}. Note that the flux limit of SKA-Low is calculated according to the SKA array for Phase I.

For an isotropic cosmic ray or neutrino flux \emph{J}(\emph{E}), during the observation time \emph{T} the number of detected events in the given energy interval between \(E_{1}\) and \(E_{2}\) can be obtained by \(N=\int_{E1}^{E2}TJ(E)A(E)dE\), where \emph{A}(\emph{E}) is the total aperture as a function of energy.
\begin{figure}[h]
\centering
\includegraphics[width=4.8in]{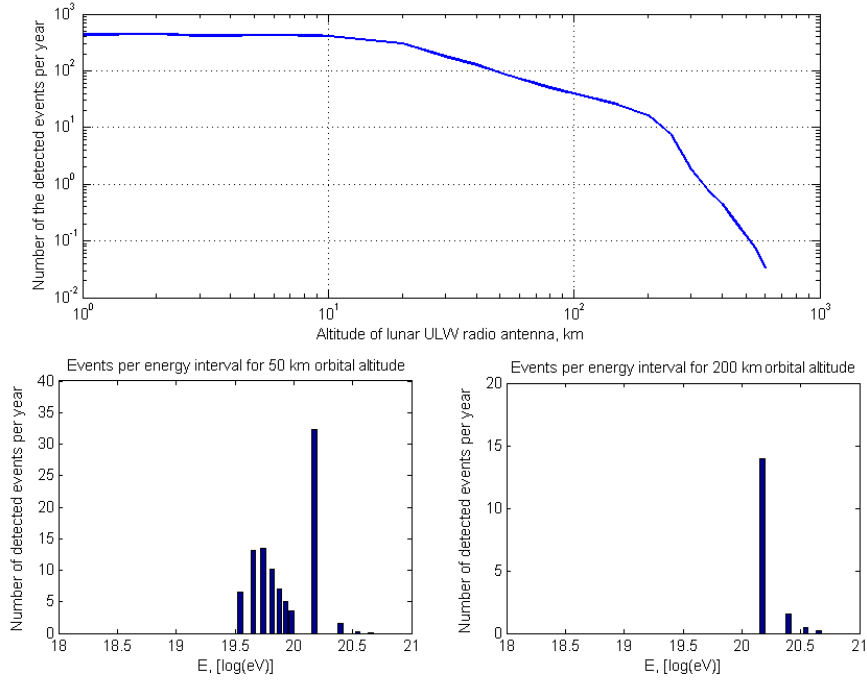}
\caption{The number of detected (per year) cosmic ray events for the energy \(E\geq 10^{18}\) eV with the flux parametrization of the Auger Collaboration \cite{aab2020}. Top, the total number detected events as a function of lunar antenna orbital altitude; Bottom, the detected events per energy interval for the lunar antenna of 50 km-high orbit (left) and 200 km-high orbit (right).
\label{fig:crevent}}
\end{figure}

Using the flux parametrization of the Auger Collaboration in Ref. \cite{aab2020}, the expected event rates of UHE cosmic rays are calculated in the energy range between \(10^{18}\) and \(10^{21}\), as shown in Figure \ref{fig:crevent}, for the lunar ULW radio antenna with different altitudes. It can be seen that the total detected cosmic ray events decreases for the antenna with higher altitudes. The detected event number reaches to about 400 for the 10 km-high orbit, and it drops rapidly from tens at the altitude of 200 km to zeros at the attitude above 500 km.  The simulation results of detected cosmic ray events per energy interval show that the events mostly lie in the higher energy range with the increasing of orbital altitude.

\subsection{UHE neutrinos}
The cascades initiated by the UHE neutrinos are different from that produced by the cosmic rays. Due to the long attenuation length for neutrinos in lunar regolith, most of the UHE neutrinos induce cascades very deep inside the Moon. Therefore, the attenuation of radio wave propagation prior to their escape from the lunar surface can not be neglected, and the contribution of both the upper and lower lunar hemisphere have to be taken into account while calculating the total aperture. Considering the neutrino absorption on the path \(L(z,\theta_{n})\), we integrate the angular aperture over z and obtain the contributions of the upper and lower hemisphere respectively \cite{gusev2006} as follows,
\begin{equation}
\label{eqn:nuaperture}
A_{\nu}(E)=A_{0}\int\Delta\Omega_{\nu}(E,\theta_{s})d\cos\theta_{s}
\end{equation}
\begin{equation}
\label{eqn:nusolidanglep}
\begin{split}
\Delta\Omega_{\nu}^{+}=\int\frac{2\ell\cos(i)}{L_{\nu N}(E_{\nu})}\ln(\frac{{\varepsilon}}{\boldsymbol{\varepsilon}_{min}}) \times\Theta[{\varepsilon}(E,\theta_{s},\theta)-{\varepsilon}_{min}]d\varphi_{n}d\cos\theta_{n}
\end{split}
\end{equation}
\begin{equation}
\label{eqn:nusolidanglen}
\begin{split}
\Delta\Omega_{\nu}^{-}=\int\frac{2\ell\cos(i)}{L_{\nu N}(E_{\nu})}\ln(\frac{{\varepsilon}}{{\varepsilon}_{min}}) \times\exp[-\frac{2R_{M}|\cos\theta_{n}|}{L_{\nu N}(E_{\nu})}]\times\Theta[{\varepsilon}(E,\theta_{s},\theta)-{\varepsilon}_{min}]d\varphi_{n}d\cos\theta_{n}
\end{split}
\end{equation}
here, it is assumed that only \(20\%\)\ of the energy of the original neutrino goes into the hadronic particle cascade. \(\ell\) is the radio wave attenuation length. Since the showers produced by the UHE neutrinos occur in the lunar soil over a large range of depths from the surface, we use the value \(\ell=29\lambda\) of lunar sub-regolith (\cite{james2009},~\cite{bray2016}) for the aperture calculation. The neutrino-nucleon interaction length was \(L_{\nu N}(\rm km)=122\rm km(E_{0}/E_{\nu})^{1/3}\), where \(E_{0}=10^{20} eV\) (\cite{gayley2009},~\cite{bray2016}). For the upper hemisphere \(\cos\theta_{n}>0\), we assume \(l/L_{\nu N}\ll1\) \cite{gusev2006}, which is valid for most of cases.
\begin{figure}[h]
\centering
\includegraphics[width=4.0in]{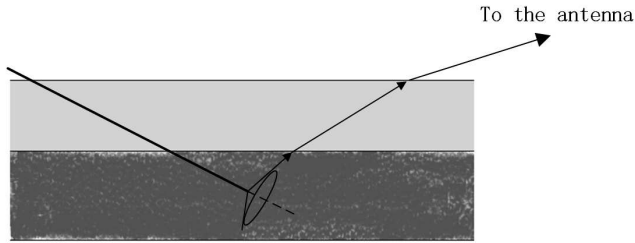}
\caption{Direct and reflected radio emissions initiated by the UHE neutrinos propagate between the lunar soil and lunar orbital radio antenna. The lunar soil is modelled with two layers, the regolith and the sub-regolith.
\label{fig:lunarsoilnu}}
\end{figure}

For the UHE neutrinos, the multi-strata structure of the lunar subsurface soil has also been taken into account in the aperture calculation. Since the cascade showers generated by the UHE neutrinos mostly occur deep inside the lunar soil, that is in the lunar sub-regolith, we consider the refractions at both the lunar surface and the subsurface strata interface, as shown in Figure \ref{fig:lunarsoilnu}. Here, the reflected signals of the Askaryan radio emissions at the lunar surface is ignored to reduce the complexity of aperture analysis.
\begin{figure}[h]
\centering
\includegraphics[width=4.8in]{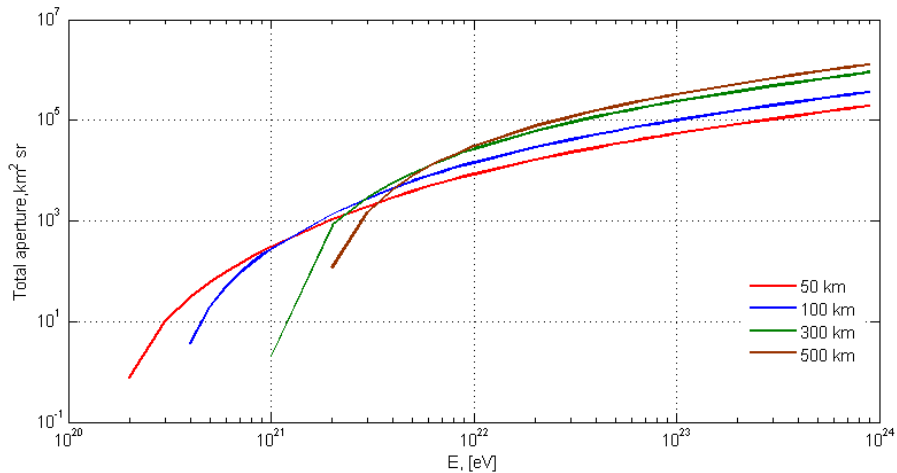}
\caption{Total aperture of the UHE neutrinos for a lunar orbital ULW radio antenna at different altitudes of 50 km, 100 km, 300 km, and 500 km with sensitivity of 1.34 uV/m/MHz.
\label{fig:nuaperture}}
\end{figure}

Similar to the cosmic rays, the total aperture of the UHE neutrinos are calculated using analytic methods for the lunar orbital ULW radio antenna at different altitudes. As shown in Figure \ref{fig:nuaperture}, the aperture increases with the rising of antenna attitude. However, the increase becomes slower above the attitude of 300 km. Being different from the cosmic rays, the aperture of the UHE neutrinos increase over the entire energy range from \(10^{20}\) eV to \(10^{24}\) eV without reaching peaks for all the four altitudes. For 50 km distance, the energy of detectable events begin at \(2\times10^{20}\) eV. With the increasing of orbital attitudes, only higher energetic neutrino events become detectable, for 500 km attitude the threshold energy of detectable events increases to \(\sim2\times10^{21}\) eV.
\begin{figure}[h]
\centering
\includegraphics[width=4.8in]{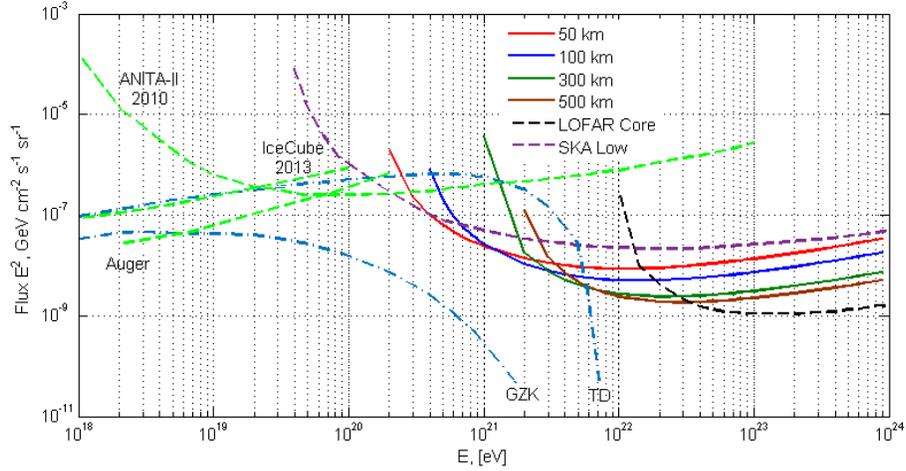}
\caption{The UHE neutrino flux of a lunar orbital ULW radio antenna at different altitudes of 50 km, 100 km, 300 km, and 500 km for one-year observation. The results are compared with the flux limits of the SKA Low for a 1000-hour observation between 100 and 350 MHz (reproduced from Bray et al., 2014), LOFAR for a 90-day observation between 125 MHz and 175 MHz (reproduced from Singh et al., 2009). The blue chain-dotted lines show the predicted fluxes from the GZK \cite{engel2001}, and topological defects (TD) \cite{semikoz2004}. The green dashed lines show the flux limits set by the past experiments of ANITA-II \cite{gorham2010}, IceCube \cite{aartsen2013} and Auger \cite{aab2015}.  \label{fig:nuflux}}
\end{figure}

Figure \ref{fig:nuflux} shows the flux limits for one-year observation with the lunar orbital antennas of different altitudes. The flux limits of the SKA-Low for a 1000-hour observation, the LOFAR for a 90-day observation, the past experiments of ANITA-II \cite{gorham2010}, IceCube \cite{aartsen2013} and Auger \cite{aab2015} are also plotted for comparison, as well as the predicted fluxes from the models of the GZK \cite{engel2001} and topological defects (TD) \cite{semikoz2004}. The results indicate that the lunar orbital observations in one year will set the comparable flux limits with the estimated fluxes of the SKA-Low and the LOFAR core (even lower than the energy of \(\sim3\times10^{22}\) eV), which signifies the higher detection efficiency of the lunar ULW radio antenna for the UHE neutrinos.

\begin{figure}[h]
\centering
\includegraphics[width=4.8in]{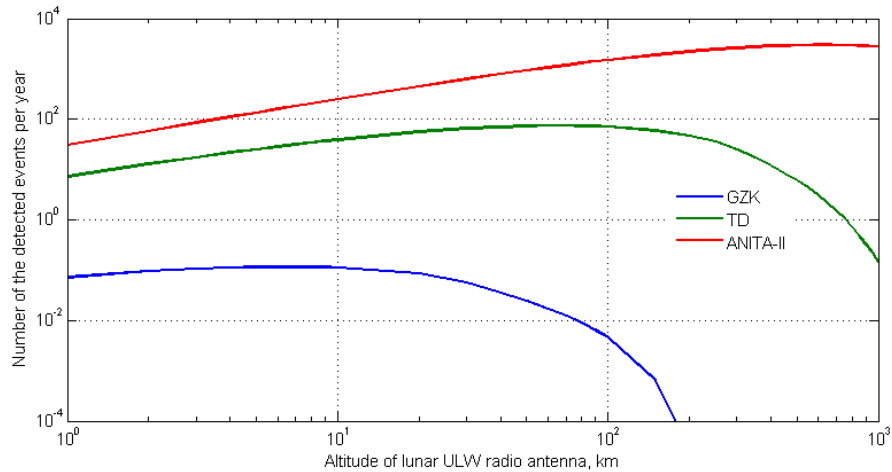}
\caption{The total number of detected (per year) neutrino events with the flux limits set by the ANITA-II experiment, and the predicted fluxes for the GZK neutrinos \cite{engel2001} and topological defects (TD) \cite{engel2001}, as a function of the lunar antenna altitude in the energy \(E\geq 10^{18}\) eV.  \label{fig:nuevent}}
\end{figure}

Based on the predicted fluxes from the models of the GZK neutrinos, topological defects (TD), the neutrino flux limit set by the experiment of ANITA-II, the expected event rates per year are estimated and presented in Figure \ref{fig:nuevent} as a function of attitudes for the lunar orbital ULW radio antenna. It is evident from Figure \ref{fig:nuevent} that the GZK neutrinos almost cannot be detected for all the altitudes. In order to increase the detection sensitivity of the GZK neutrinos, the aperture must be improved by increasing the effective receiving area of the radio telescope, for instance, employing a radio array. For the TD neutrino detection, the event rate depends on the altitude, and reaches the peak of about 70 per year at the attitude of about 100 km, which could be the optimum attitude for a lunar orbital ULW antenna. With the neutrino flux limit of ANITA-II, the expected neutrino events increase exponentially at the altitude range from 1 km to 1000 km, and thousands of events per year could be detected at most by the lunar ULW antenna.

\section{Characteristic analysis of UHE cosmic rays and neutrinos}
\label{sec:characteristic}
For the detection of the UHE cosmic rays and neutrinos, the ultimate goal is to reconstruct the events to obtain all the parameters characterizing the cosmic rays or neutrinos, including the source, the primary particle energy, etc. In the literatures (\cite{gusev2015},~\cite{gusev2017}), a random search method was employed to determine the UHECR event parameters for the LORD experiment under the assumption that the cosmic ray flux is known. Due to the lack of direct information, the reconstructed parameters have large uncertainties even for the detection with two satellites. Compared with the LORD experiment, the future lunar ULW radio missions like DSL (\cite{boonstra2016},~\cite{chen2021}), OLFAR \cite{engelen2011} have more advantages in the UHE particle detection. They consist of many antennas onboard micro-satellites to form a large radio array, which make it feasible to detect the UHECR and UHEC\(\nu\) with multi satellites simultaneously. Furthermore, in lunar ULW radio observations, the tripole antenna can measure the three-dimension electric fields, which make it possible to measure the three-dimension polarization of the Askaryan radio emission.
\begin{figure}[h]
\centering
\includegraphics[width=4.8in]{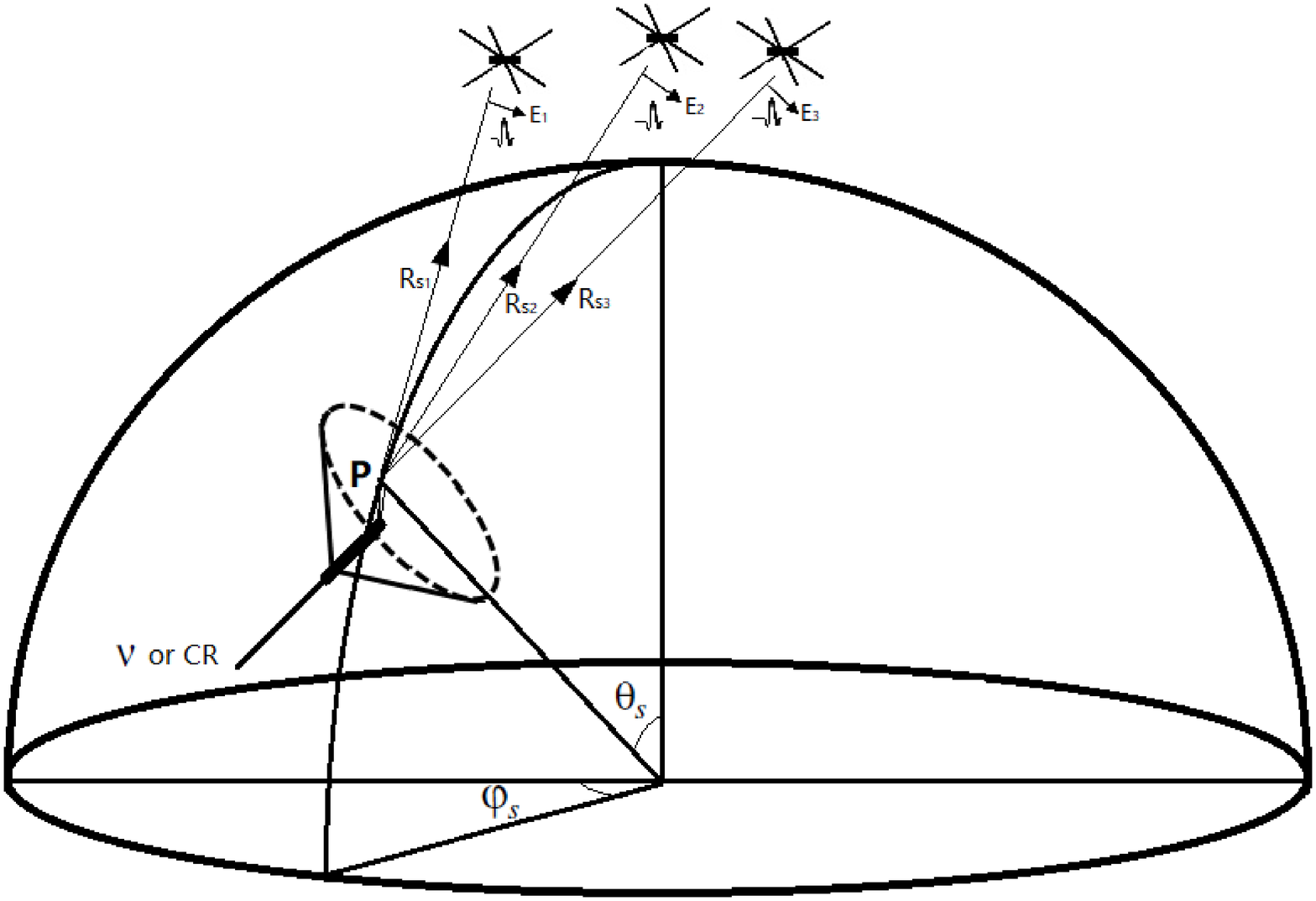}
\caption{Scheme of Askaryan pulse detections in the lunar soil with three antennas onboard micro-satellites.  \label{fig:multidetections}}
\end{figure}

In the detection of the UHECR and UHEC\(\nu\), if the Askaryan radio pulses can be observed by multi antennas onboard lunar satellites, the time delays can be measured between the radio pulses sensed by different antennas, and the particle cascade location (\textbf{P}) on the lunar surface can be solely determined with the time delays among at least three non-linearly distributed antennas, as shown in Figure \ref{fig:multidetections}. We know the Askaryan radio emission is highly linearly polarized, which is always in the same plane of the Poynting vector and the shower axis (\cite{james2019},~\cite{heino2004},~\cite{gorham2004}). Therefore, using the three-dimension polarizations of the electric fields (\(\rm E_{1},\rm E_{2}, \rm E_{3}\)) measured by the tripole antennas onboard three orbital satellites, the show axis can be determined by the intersection line of the three planes of \(\rm E_{1}-R_{\rm s1},\rm E_{2}-R_{\rm s2}, \rm E_{3}-R_{\rm s3}\) shown in Figure \ref{fig:multidetections}, which means the original direction of the UHECR or UHEC\(\nu\) is known now. For a linear ULW radio array like DSL, it is necessary to note that the cascade location deduced by the time delays could have two candidates symmetric with respect to the satellite orbit plane. However, since only the right location can account with the fact that the shower axis, the Poynting vectors and the polarizations of the Askaryan radio emissions are in the same plane, the measured polarizations of the coherent radio pulses on three tripole antennas will definitely exclude one of the candidate locations except for the case that the shower axis is parallel to the satellite orbit plane.

Once the shower axis and the location of the UHE particle cascade are confirmed, the distance \(R_{\rm s}\) between the particle cascade and the radio detector, the three angles \(\theta_{n}, r, i\) shown in Figure \ref{fig:scheme} will also be known according to the simple geometrical relationship. Using the formulae (\ref{eqn:intensity}),(\ref{eqn:angular}) and (\ref{eqn:transcoeff}), the particle energy \emph{E} can be calculated with the intensities of electric fields induced on the lunar orbital antennas. The results of the three satellites will be cross-checked with each other to improve the accuracy and reliability.

\section{Summary and conclusions}
\label{sec:conclusion}
In the present study, using the lunar Askaryan technique, the detection of the UHE cosmic rays and neutrinos has been analytically analyzed for the future lunar ULW radio missions. Our results indicate that the single lunar ULW radio antenna could detect the cosmic ray and neutrino flux as low as that observed by the present or future experiments in the energy range \(E > 10^{20} \rm eV\). With the known flux of the UHE cosmic rays, the simulated detectable events per year has found to increase when the antenna altitude is lowered, and it reaches the peak at the altitude of \(\sim10\) km. For the UHE neutrinos, during the observations in one year, it is shown that the single lunar ULW radio antenna could detect \(\sim70\) topological defect neutrinos at an optimal altitude of 100 km, while the GZK neutrinos could not be detected.

Investigating the properties of the lunar Cherenkov radiation, it has been demonstrated that the UHECR or UHEC\(\nu\) events could be reconstructed directly using the radio observations with at least three tripole antennas onboard the lunar satellites that makes it feasible to detect the UHECR and UHEC\(\nu\) with the lunar ULW radio mission. The method could also be used to build a dedicated lunar radio telescope for the detections of the UHE cosmic ray and the UHE neutrinos.

\section*{Acknowledgements}
This work is funded by the National Natural Science Foundation of China (NSFC) under No.11941003,11790305, 11573043, the CE-4 mission of the Chinese Lunar Exploration Program: the Netherlands-China Low Frequency Explorer (NCLE), and the Chinese Academy of Science Strategic Priority Research Program XDA15020200. The authors would like to thank Dr. C.W. James from the University of Adelaide, Australia, and Prof. Dr. Xuelei Chen from National Astronomical Observatories, Chinese Academy of Sciences for all the useful discussions. In addition the authors thank the referees for their useful comments and suggestions.

\begin{appendix}
\renewcommand{\thesection}{Appendix}
\setcounter{equation}{0}
\renewcommand\theequation{A.\arabic{equation}}
\section{A. Aperture calculation of UHE cosmic rays and neutrinos}
Based on the analytic methods used in \cite{gusev2006}, the modifications have been made to the aperture analysis for both cosmic rays and neutrinos in this paper. The differences include the calculations of electric field intensity and angular spread, the attenuation length and neutrino-nucleon interaction length of lunar regolith, the model of antenna sensitivity, etc., which have been addressed in this paper.

\begin{figure}[h]
\centering
\includegraphics[width=\linewidth]{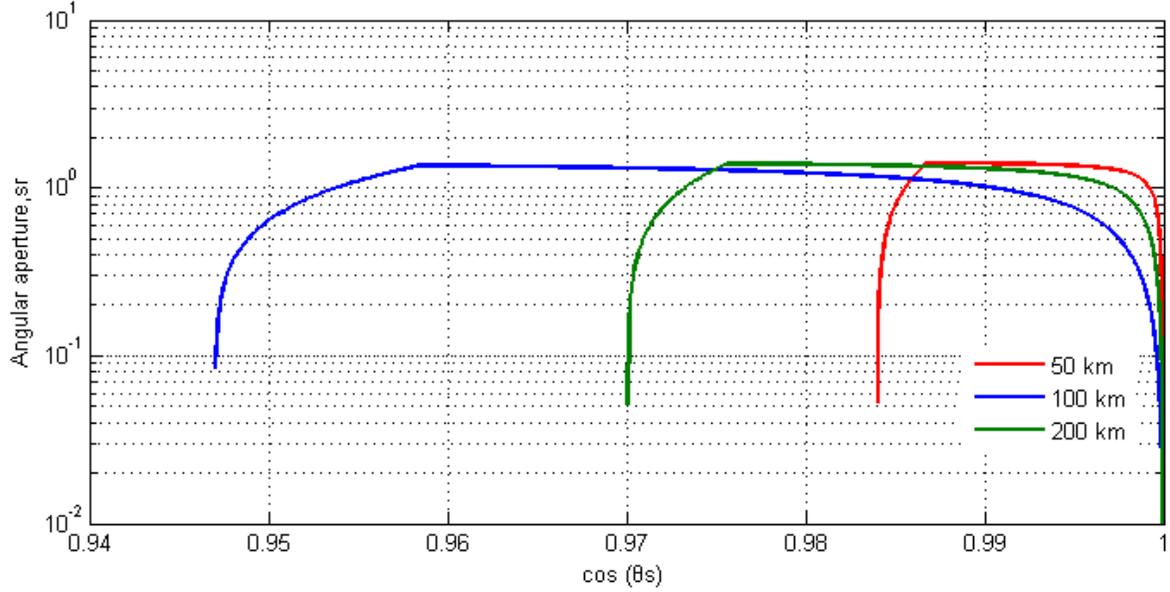}
\caption{Angular aperture \(\Delta\Omega_{CR}\) for the cosmic ray detection with a lunar orbital ULW radio antenna at different altitudes of 50km, 100km and 200km, and for the initial cosmic ray energy \(W=10^{21} eV\).
\label{fig:crangular}}
\end{figure}

For the detection of UHE cosmic rays, the total aperture is calculated by integrating the angular aperture over the upper hemisphere surface, where
\begin{equation}
\Delta\Omega_{CR}=\int_{0}^{1}\cos\theta_{n}d\cos\theta_{n}\int_{-\pi}^{\pi}\Theta [\boldsymbol{\varepsilon}(E,\theta_{s},\theta)-\boldsymbol{\varepsilon_{\rm rms}}]d\varphi_{n}
\end{equation}
In the angular coordinate (\(\theta,\varphi\)) with a polar axis aligned with the radiation direction (see Figure \ref{fig:scheme}), it can be derived as,
\begin{equation}
\cos\theta_{n}=\sin\theta\sin i\cos\varphi-\cos\theta\cos i
\end{equation}
\begin{equation}
d\Omega=-d\cos\theta_{n}d\varphi_{n}=-d\cos\theta d\varphi\
\end{equation}
Then the angular aperture can be re-written as,
\begin{equation}
\begin{split}
\Delta\Omega_{CR}=2\int_{\cos\theta_{max}}^{\cos\theta_{min}}d\cos\theta\times\int_{0}^{\varphi_{m}}(\sin\theta\sin i\cos\varphi-\cos\theta\cos i)d\varphi
\end{split}
\end{equation}
here \(\cos\varphi_{m}=(\cos\theta\cos i)/(\sin\theta\sin i)\). \(\theta_{min}\) and \(\theta_{max}\) are the minimum and maximum angels of incidence within which the detected radiation intensity \(\boldsymbol{\varepsilon}\) is stronger than \(\boldsymbol{\varepsilon_{min}}\). The angular apertures calculated as functions of \(\cos\theta_{s}\) are shown in Figure \ref{fig:crangular} .

For the detection of UHE neutrinos, the neutrino absorption on the path up to the shower production point has to be taken into account in the calculation of angular aperture, as well as the absorption of radio wave in the lunar regolith.
\begin{equation}
\begin{split}
\Delta\Omega_{\nu}=\int\frac{dz}{L_{\nu N}}\int \Theta\{\varepsilon(E,\theta_{s},\theta)\times\exp[-z/(2\ell\cos i)]-\varepsilon_{\rm rms}\}\times\exp[-L(z,\theta_{n})/L_{\nu N}(E_{\nu})]
\end{split}
\end{equation}
here, \(L(z,\theta_{n})\approx z/\cos\theta_{n}\) for the lunar upper hemisphere \(\cos \theta_{n}>0\), and \(L(z,\theta_{n})\approx 2R_{\rm M}\mid\cos\theta_{n}\mid\) for the lunar lower hemisphere \(\cos \theta_{n}<0\). By integrating over the cascade depth \emph{z}, we obtain, respectively, the contributions of the upper and lower hemispheres to the neutrino angular aperture.
\begin{equation}
\label{eqn:nusolidangleps}
\begin{split}
\Delta\Omega_{\nu}^{+}\simeq\int_{0}^{z_{max}}\frac{1}{L_{\nu N}}\exp(-\frac{z}{L_{\nu N}\cos\theta_{n}})dz\times\int\Theta[{\varepsilon}(E,\theta_{s},\theta)-{\varepsilon}_{min}] d\varphi_{n}d\cos\theta_{n} \\
\simeq\int\cos\theta_{n}\{1-\exp[\frac{2\ell\cos i}{L_{\nu N}\cos\theta_{n}}\ln(\frac{\varepsilon}{\varepsilon_{min}})]\}
\times\Theta[{\varepsilon}(E,\theta_{s},\theta)-{\varepsilon}_{min}] d\varphi_{n}d\cos\theta_{n}
\end{split}
\end{equation}
\begin{equation}
\begin{split}
\Delta\Omega_{\nu}^{-}\simeq\int_{0}^{z_{max}}\frac{1}{L_{\nu N}}\exp(-\frac{2R_{\rm M}\cos\theta_{n}}{L_{\nu N}})dz \times\int\Theta[{\varepsilon}(E,\theta_{s},\theta)-{\varepsilon}_{min}] d\varphi_{n}d\cos\theta_{n} \\
\simeq\int\frac{2\ell\cos i}{L_{\nu N}}\ln(\frac{\varepsilon}{\varepsilon_{min}})\exp(-\frac{2R_{\rm M}\cos\theta_{n}}{L_{\nu N}})\times\Theta[{\varepsilon}(E,\theta_{s},\theta)-{\varepsilon}_{min}] d\varphi_{n}d\cos\theta_{n}
\end{split}
\end{equation}
where, \(z_{max}=2\ell\cos i\ln(\varepsilon/\varepsilon_{min})\). For the ultra-high energy neutrino, when \(\ell/L_{\nu N}\ll1\), formula \ref{eqn:nusolidangleps} will be simplified to the formula (\ref{eqn:nusolidanglep}). The angular aperture is then calculated numerically.
\end{appendix}

\bibliographystyle{spbasic}      


\label{lastpage}

\end{document}